\documentclass[twocolumn]{aastex631}

\newcommand{\A}{\text{\normalfont\AA}}

\def \ebmv{E(B-V)}

\def \halpha{H$\alpha$}

\def \h2{{\rm H_{2}}}

\def \halpha{H$\alpha$}
\def \cii{\ion{C}{2}}
\def \oii{[\ion{O}{2}]}
\def \oiii{[\ion{O}{3}]}
\def \hii{\ion{H}{2}}

\def \dn4000{D_{{\rm n}}(4000) }

\def \name{JADES-GS-z7-01-QU}
\def \sphinxtwenty{\textsc{Sphinx$^{\rm 20}$}}
\def \sphinx{\textsc{Sphinx}}

\begin{document}

\title{Dead or Alive? How Bursty Star Formation and\\
Patchy Dust Can Cause Temporary Quiescence in High Redshift Galaxies}

\author[0000-0002-9382-9832]{Andreas L. Faisst}
\affiliation{Caltech/IPAC, 1200 E. California Blvd. Pasadena, CA 91125, USA}

\author[0000-0002-8512-1404]{Takahiro Morishita}
\affiliation{Caltech/IPAC, 1200 E. California Blvd. Pasadena, CA 91125, USA}



\begin{abstract}

The recent discovery of a galaxy at $z=7.3$ with undetected optical emission lines and a blue UV to optical continuum ratio in JWST spectroscopy is surprising and needs to be explained physically. Here, we explore two possibilities that could cause such a seemingly quiescent $\sim5\times10^8\,{\rm M_\odot}$ galaxy in the early Universe: {\it (i)} stochastic variations in the star formation history (SFH) and {\it (ii)} the effect of spatially varying dust attenuation on the measured line and continuum emission properties. Both scenarios can play at the same time to amplify the effect.
A stochastic star formation model (similar to realistic SFHs from hydrodynamical simulations of similar-mass galaxies) can create such observed properties if star formation is fast-varying with a correlation time of $<150\,{\rm Myrs}$ given a reasonable burst amplitude of $\sim 0.6\,{\rm dex}$. The total time spent in this state is less than $20\,{\rm Myrs}$, and the likelihood of such a state to occur over $500\,{\rm Myrs}$ at $z=7$ is $\sim50\%$ (consistent with current observations).
On the other hand, we show that a spectrum with blue UV continuum and lack of emission lines can be reproduced by a blue$+$red composite spectrum. The UV continuum is emitted from dust-free density bounded \hii~regions (blue component), while the red component is a dust-obscured starburst with weakened emission lines due to strong differential dust attenuation between stellar and nebular emission.
Future resolving far-infrared observations with ALMA will shed light on the latter scenario.

\end{abstract}



\section{Introduction} \label{sec:intro}

Galaxies spend a significant amount of their lives on the star forming main-sequence \citep{noeske07,daddi07}, a tight $\sim0.3\,{\rm dex}$ relation between stellar mass and star formation rate (SFR). This relation is maintained by an equilibrium between gas inflow, gas consumption through star formation, and gas removal through outflow (e.g., stellar winds, Supernovae, or Active Galactic Nuclei [AGN]), as explained by simple analytical models \citep[e.g.,][]{dave12,lilly13,feldmann15}.
After the peak of cosmic star formation rate density at $z\sim2$ \citep[e.g.,][]{madau14} an increasing number of quiescent galaxies is observed, which deviate from the star-forming main sequence by a significantly reduced star formation \citep[e.g.,][]{ilbert10,abraham07,bell04}. Quiescent galaxies are mainly characterized by their red colors and strong Balmer or $4000\,{\rm \AA}$ breaks due to 50 Myr or older populations as well as the lack of emission lines signaling the lack of recent star formation.
More quantitatively, the state of quiescence can be defined as a galaxy having a specific star formation rate (sSFR) below a certain threshold. It is found that a sSFR threshold of $0.2/t_{\rm univ}(z)\,{\rm Gyr^{-1}}$ (where $t_{\rm univ}(z)$ is the age of the Universe at redshift $z$) is consistent with various color selections \cite[see][]{pacifici16,faisst17}.
The probability of being quiescent depends on the environment and stellar (halo) mass of a galaxy \citep[e.g.,][]{peng10}. Quenching can happen through many avenues, for example by gas shock-heating in the most massive dark matter halos \citep[common at high masses;][]{croton06,cattaneo08} and interactions with starbursts or feedback from AGN and Supernovae \citep[also common at lower masses;][]{toomre72,haas13,schawinski14}.

While the quiescent galaxy population is established at $z<2$, there are open questions regarding their progenitors and when and how the first quiescent galaxies emerge.
With the advent of sensitive infrared detectors and surely the launch of the {\it James Webb Space Telescope} (JWST), it became feasible to spectroscopically confirm quiescent galaxies at to Epoch of Reionization at $z=6$ \citep[][]{glazebrook17,schreiber18,tanaka19,forrest20,valentino20,eugenio21,kubo21,nanayakkara22,carnall23a,carnall23b,valentino23,looser23}. 
Most of these findings are massive galaxies ($\gtrsim 10^{10}\,{\rm M_\odot}$) that are thought to be quiescent for many $100\,{\rm Myrs}$ to a Gyr. However, the fate of less massive quiescent galaxies is unclear because their star formation is suggested to significantly oscillate around the equilibrium state ({\it i.e.} the main sequence) due to the continuous adjusting between gas supply and consumption, hence leading to ``temporary quiescence'' \citep[e.g.,][]{tacchella16}. It is also suggested by simulations \citep[e.g.,][]{remus23} that quiescent galaxies can be rejuvenated leading to star formation in the outer disk.

An interesting exception is the finding of a recently quenched galaxy (in the following {\it \name}) at a spectroscopic redshift of $z=7.3$ reported by \citet{looser23}.
The galaxy is curious as it is lacking optical emission lines and at the same time shows a blue UV to optical flux ratio in the present JWST/NIRSpec spectra.
A quiescent galaxy at such a high redshift (where commonly a significant amount of molecular gas is available) may be surprising. Mass quenching is likely not applicable due to the lower masses of dark-matter halos at those epochs, instead feedback or variations in the star formation history (SFH) of the galaxy are more likely.
The lack of optical line emission suggests the lack star formation in the past $\sim10\,{\rm Myrs}$, indicative of a short-term variation in the SFH of this galaxy \citep[e.g.,][]{kennicutt98}. Furthermore, the galaxy shows a blue UV to optical flux ratio of $\sim 3$ (in $\lambda f_{\rm \lambda}$), indicative of young stars and/or low stellar continuum dust obscuration. Since the galaxy is at a relatively low stellar mass ($\sim 5\times 10^8\,{\rm M_\odot}$), the most likely scenarios involve temporal quenching by star formation or AGN feedback \citep[e.g.,][]{gelli23,lovell23} or galaxy-galaxy interactions \citep{asada23}. A recent study found that the most dominant quenching mechanisms for low-mass high-redshift galaxies may be accretion disk feedback triggered by disk instabilities \citep{xie24}.
Alternatively, an unlucky geometric distribution of dust along the line of sight could attenuate the nebular emission and mimic quiescence.
In the meantime, similar galaxies have been found at $z\sim4.5$ \citep{looser23b} and some that are on the way to a quiescent state at $z\sim5.2$ \citep{strait23}, in agreement with the bursty star formation expected at these redshifts \citep[e.g.,][]{faisst19}.

\begin{figure*}[t!]
\centering
\includegraphics[angle=0,width=1.0\columnwidth]{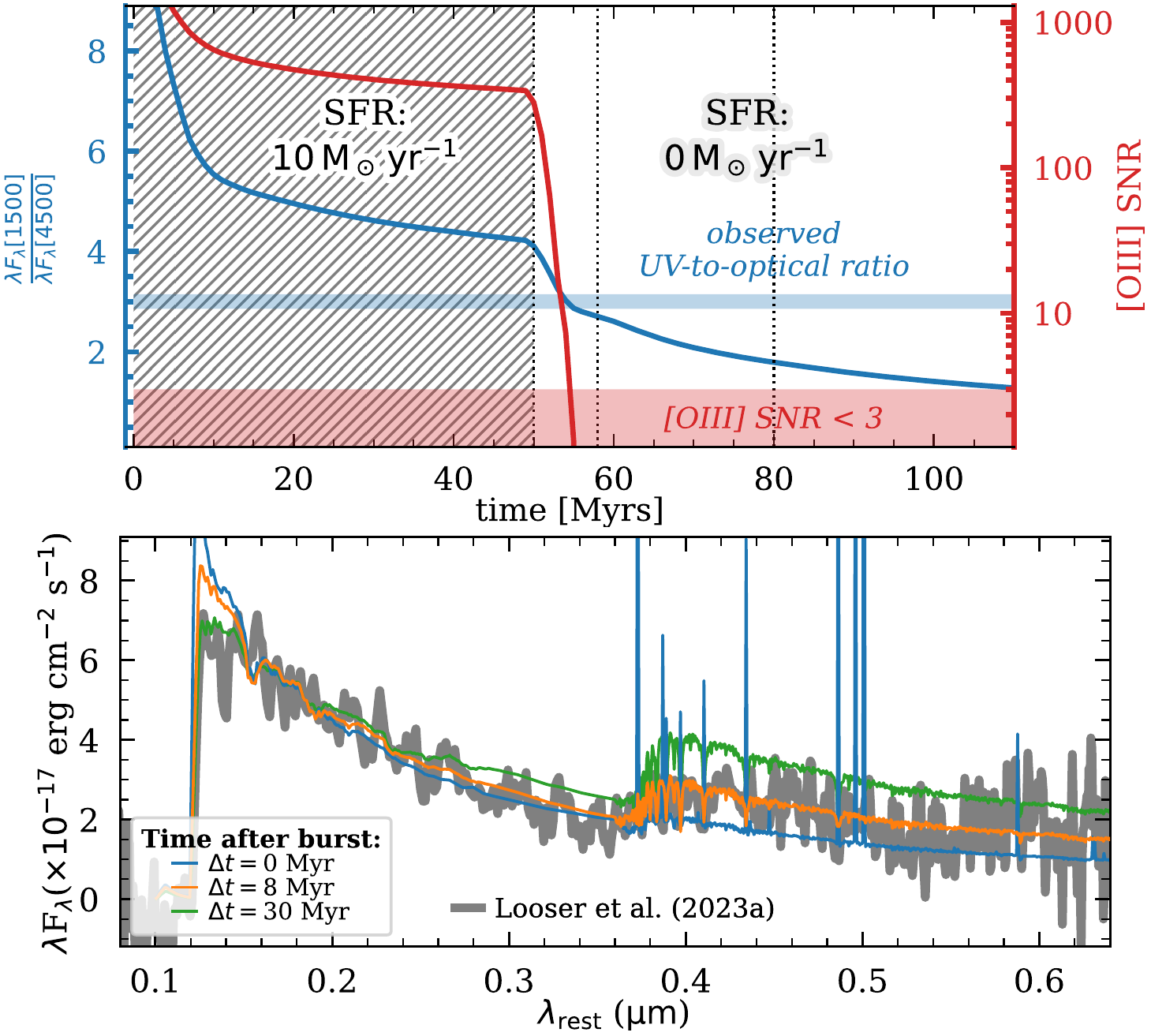}
\includegraphics[angle=0,width=1.0\columnwidth]{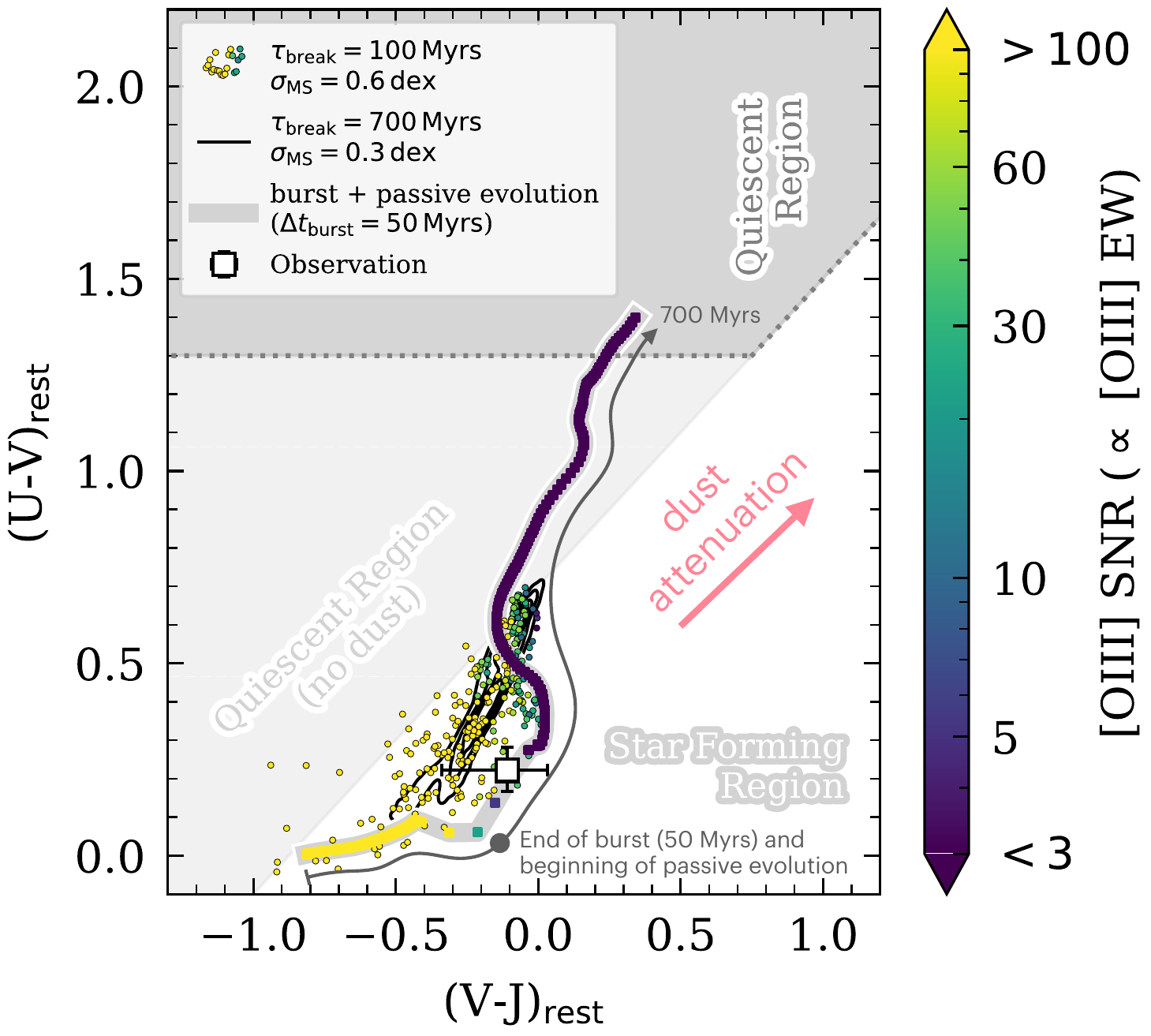}
\caption{{\it Top-Left:} UV-to-optical flux ratio (blue) and \oiii~SNR (red) as a function of time for a simple burst model with $10\,{\rm M_\odot\,yr^{-1}}$ over $50\,{\rm Myrs}$ (hatched area). The observed values for \name~are indicated as blue and red horizontal filled areas, respectively.
{\it Bottom-Left:} Comparison of model spectra at different times after the burst (indicated by dashed vertical lines in top-left panel) with the observed spectrum of \name~from \citet{looser23}.
{\it Right:} UVJ diagram showing the observe colors of \name~(square) and different models. The simplistic burst$+$passive evolution model with maximal age of $700\,{\rm Myrs}$ is shown as thick line with squares colored by the \oiii~SNR (proportional to the \oiii~equivalent width). The colored points show a stochastic model ($\tau_{\rm break}=100\,{\rm Myrs}$, $\sigma_{\rm MS} = 0.6\,{\rm dex}$) over $700\,{\rm Myrs}$. The background contours show a stochastic model with $\tau_{\rm break}=700\,{\rm Myrs}$ and $\sigma_{\rm MS} = 0.3\,{\rm dex}$ resulting in less scatter. Note that dust changes the colors long the separation line between star-forming and quiescence (gray area). No dust is included in this model, therefore in the case of the burst$+$passive model the points do not land in the original UVJ passive region (dark gray). In reality, they would be pushed to the upper right into that passive region due to dust attenuation.
\label{fig:uvj_model}}
\end{figure*}

Several studies tried to explain the nature of \name, specifically the combination of a blue UV to optical flux ratio and the absence of optical emission lines. Simulations can explain quiescent galaxies and their number densities in the Epoch of Reionization \citep[e.g., ][]{dome23,vikaeus23,lovell23,gelli23,pallottini23}, and temporal quiescence is also expected given the burstiness of star formation especially at lower stellar masses at high redshifts \citep[e.g.,][]{dome23,sun23a,sun23b,abramson20,faisst19,sparre17,kelson14}.
However, the exact spectral energy distribution (SED) of JADES-GS-z7-01-QU could not be reproduced so far \citep[see, e.g.,][]{gelli23}.

In this study, we aim at explaining the observation of a galaxy such as \name~by two (not mutually exclusive) ideas: {\it (i)} the stochastic variations in the SFH of high-$z$ galaxies (Section~\ref{sec:stoch}), and {\it (ii)} spatial variations in the dust attenuation (Section~\ref{sec:dust}).
We assume a $\Lambda$CDM cosmology with $H_0 = 70\,{\rm km\,s^{-1}\,Mpc^{-1}}$, $\Omega_\Lambda = 0.7$, and $\Omega_{\rm m} = 0.3$. Magnitudes are given in the AB system \citep{oke74}, and stellar masses and SFRs are normalized to a \citet{chabrier03} initial mass function (IMF).

\section{Quiescence in Stochastic Star Formation Histories}\label{sec:stoch}

In this section, we study the properties of \name~under the assumption that the galaxy follows a stochastic SFHs. We want to recover the color properties as well as the lack of emission lines observed with JWST NIRCam and NIRSpec data.
Specifically, we observe two conditions \citep{looser23}: 

\paragraph{{\bf Condition 1.}~}~A conservative non-detection of optical emission lines, specifically \oii$_{\rm \lambda 3727}$~and \oiii$_{\rm \lambda 5007}$ at no more than $3\sigma$ assuming a continuum signal-to-noise ratio (SNR) of $5$.

\paragraph{{\bf Condition 2.}~}~A blue UV to optical $\lambda f_{\rm \lambda}$ flux ratio of $3\pm0.3$ as observed.\vspace{3mm}

In the following, we show that these observed properties of \name~can be reproduced after a starburst. We then move on to estimate the probability of such a scenario to happen within more complicated stochastic representations of SFHs.

\subsection{Single Starburst}\label{sec:burst}

The lack of optical emission lines is indicative of a lack of star formation in the past $\sim3-10\,{\rm Myrs}$. This condition can be reproduced by a single starburst (Figure~\ref{fig:uvj_model}).
Specifically, we assume a burst with a SFR of $10\,{\rm M_\odot\,yr^{-1}}$ and a length of $50\,{\rm Myrs}$ (indicated by the black hatched area). We assume that all mass is produced during that time, while any previous star formation only has negligible impacts on the results observed here. Furthermore, we do not assume dust attenuation in this simple model.
The blue and red lines on the top-left panel in Figure~\ref{fig:uvj_model} show the UV-to-optical flux ratio (condition 2) and the \oiii~SNR (condition 1) in this scenario, respectively. The observed values are indicated as blue and red horizontal areas. Shortly ($\sim5\,{\rm Myrs}$) after the end of the burst, both conditions 1 and 2 are met. We note that the sharp drop in the \oiii~line emission is due to the lack of ionizing flux of intermediate-age stars.
The bottom-left panel shows a comparison of the model spectra at three different times after the burst to the observed spectrum in \citet{looser23}.
The right panel of Figure~\ref{fig:uvj_model} shows the burst model with subsequent passive evolution (line) on the UVJ diagram. The corresponding \oiii~SNR (proportional to the \oiii~equivalent width) is shown in color and it drops to zero after the burst. This diagram is commonly used to separate star-forming from quiescent galaxies \citep[e.g.,][]{wuyts07}. Assuming passive evolution, a galaxy would slowly evolve towards the quiescent regime (here maximum at $t_{\rm univ}(z=7.3) \sim 700\,{\rm Myrs}$). Note that dust attenuation moves colors to the upper right (red arrow), parallel to the dividing line between star formation and quiescence, hence this analysis is insensitive to the assumed dust attenuation. The observed colors of \name~are shown by the large square, and they agree well with the burst$+$passive evolution model right at the end of the burst.

The above analysis shows that the conditions of \name~can be explained right after a burst of star formation, and it assumes that the galaxy stays quiescent afterwards. It is suggested that star formation can be quenched for good in low-mass dark matter halos during the epoch of reionization, giving rise to quiescent dwarf galaxies in the present days \citep[e.g.,][]{bullock00,brown14}. However, realistically, the SFH of galaxies at these masses is much more complex and a pure passive evolution after a burst is unlikely given the large gas abundances at those early times (however, see \citealt{xie24} for theoretical evidence that most early quiescent galaxies stay quiescent for $\sim1\,{\rm Gyr}$). Dark matter halo related quenching is not effective at low masses.
The question of how often such a scenario would occur in a more realistic stochastic SFH therefore arises. This is explored in the next sections.

\subsection{Stochastic Star Formation Histories}\label{sec:stocSFH}

\subsubsection{Synthetic spectra derived from stochastic SFHs}\label{sec:syntspec}

The SFH of galaxies can be characterized by stochastic processes depending on feedback, inflow, and outflow of gas. This stochasticity will lead to starbursts and short periods of quiescence, potentially giving rise to \name~as shown in Section~\ref{sec:burst}. Here, we use an approach similar to \citet{caplar19} to create model SFHs with different statistical properties.
Specifically, we use the \textsc{DELightcurveSimulation} Python package\footnote{\url{https://github.com/samconnolly/DELightcurveSimulation}} \citep{connolly16}, which is based on the \citet{emmanoulopoulos13} light curve simulation algorithm. In the following, we assume a generalized power spectrum density (PSD) based on a damped random walk
\begin{equation}
    {\rm PSD}(f) = \frac{\sigma^2}{1 + (\tau_{\rm break}\,f)^\alpha},
\end{equation}
where $f$ is the frequency ({\it i.e.} inverse time in Myrs), $\tau_{\rm break}$ is related to the decorrelation time scale of star formation over time, $\sigma$ is the long-term variability (or amplitude), and $\alpha$ describes the slope of the PSD ($\alpha=2$ corresponds to a damped random walk). Note that in the following $\sigma = \sigma_{\rm MS}$, the amplitude of star formation burstiness with respect to the main-sequence in dex, is assumed. SFHs are created by using the {\it Simulate\_TK\_Lightcurve} function using the \citet{timmer95} method and the above parameterization of the PSD.
Synthetic spectra are derived using the Flexible Stellar Population Synthesis \citep[FSPS; ][]{conroy09,conroy10} code\footnote{We are using the Python implementation, \url{https://dfm.io/python-fsps/} \citep{johnson23}.} by feeding it the stochastic SFHs obtained before. We enable emission lines (with $\log U = -2$), assume a $10-50\%$ solar metallicity of gas and stars (the exact value does not impact the following results), a \citet{chabrier03} IMF, and ``Padova 1994'' stellar evolutionary tracks \citep{bertelli94} together with a Basel Stellar Libary \citep[BaSeL, version 3.1; ][]{lejeune97,lejeune98,westera03} complemented by the empirical \textsc{STELIB} library \citep{leborgne03}.
We do not include additional ``birth clouds'' or diffuse dust emission in this model. Instead, we will add the effects of dust later when we extract the fluxes of the UV continuum and optical emission lines.

\subsubsection{How likely is it to observe \name?}\label{sec:occurrence}
For deriving a probability to observe a galaxy with similar properties as \name~within a given time frame, we create several 1000 synthetic spectra for various $\tau_{\rm break}$ from 50 to 500 Myrs and $\sigma~=~0.4,0.6,0.8\,{\rm dex}$.
Furthermore, we fix $\alpha=2$ (see \citealt{caplar19} for observational evidence) and the redshift is assumed to be $z=7.3$. These spectra are then evaluated under the two conditions outlined above. If both conditions are satisfied, a suitable analog of \name~has been found.

\begin{figure}[t!]
\centering
\includegraphics[angle=0,width=1\columnwidth]{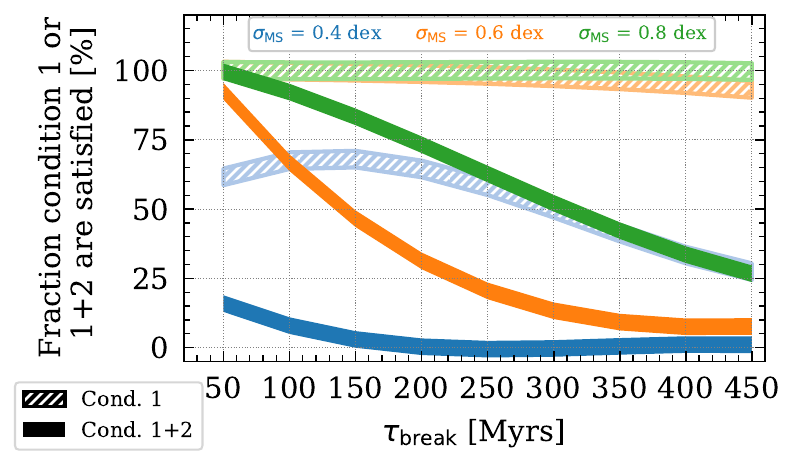}\\
\includegraphics[angle=0,width=1\columnwidth]{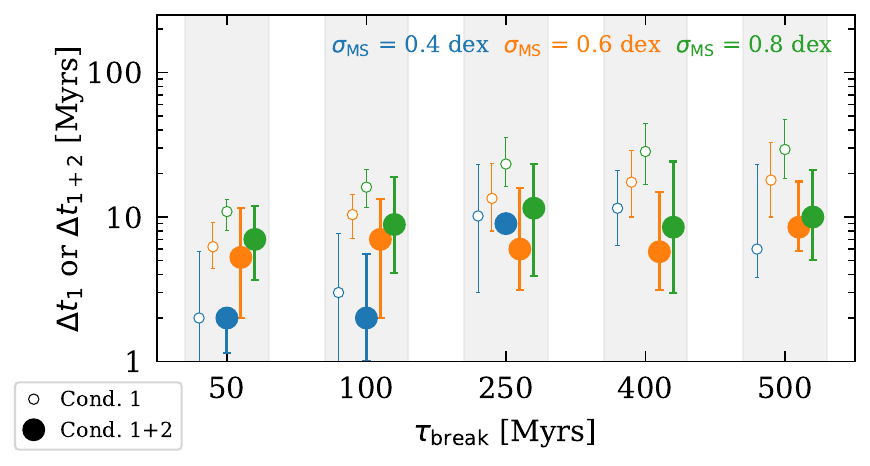}
\caption{{\it Top:} Probability (in per-cent) of the occurrence of a galaxy with condition 1 (no emission lines; hatched) or conditions $1+2$ (no emission lines and blue UV to optical flux ratio; filled) within $500\,{\rm Myrs}$ at $z=7.3$ for different main-sequence scatter amplitudes ($\sigma_{\rm MS}$) and $\tau_{\rm break}$. The calculation is based on synthetic spectra derived for stochastic SFHs (Section~\ref{sec:syntspec}).
{\it Bottom:} Time spend in either condition 1 (small empty circles) or conditions $1+2$ (large filled circles) for the same selection of $\sigma_{\rm MS}$ and $\tau_{\rm break}$. The symbols are offset in $x-$direction to improve visual clarity.
\label{fig:stochastic}}
\end{figure}

The number of $30\,{\rm Myr}$ time-slices that meet either condition 1 or both conditions $1+2$ are then computed within a $500\,{\rm Myr}$ time window to derive a probability of such a galaxy to occur over that time frame. The top panel in Figure~\ref{fig:stochastic} summarizes this likelihood in per-cent for the different $\tau_{\rm break}$ and $\sigma_{\rm MS}$. Generally, we find that the likelihood to observe either condition {\it increases} with larger amplitude ($\sigma_{\rm MS}$), and {\it decreases} with increasing $\tau_{\rm break}$. This is not surprising $-$ high amplitudes are necessary to produce a blue UV continuum and short correlation timescales are needed to produce frequent fast drops in star formation on the timescale of the de-excitation of nebular emission lines (c.f. Section~\ref{sec:burst}).
Naturally, observing both a lack of emission lines {\it and} a blue UV to optical flux ratio is significantly less likely, which is the main difficulty in reconciling the observed properties of \name~\citep[see][]{gelli23}.
We note that in the case of this analysis, $\tau_{\rm break}$ and $\sigma_{\rm MS}$ are degenerate. We can break this degeneracy by fixing $\sigma_{\rm MS}$, {\it i.e.} the burstiness of the star formation, which is a quantity that can be measured from observations.
Assuming a fiducial $\sigma_{\rm MS} = 0.5-0.6\,{\rm dex}$ \citep[which is likely a good assumption given the measured burstiness of low-mass galaxies at high redshifts, see][]{faisst19,looser23,popesso23,mehta23,cole23} the occurrence of a galaxy with properties similar to \name~is most likely ($>50\%$) within $500\,{\rm Myrs}$ for a considerably varying SFH at a rate of $\tau_{\rm break} \lesssim 150\,{\rm Myrs}$.

The bottom panel of Figure~\ref{fig:stochastic} shows the estimated time a galaxy would spend on average in either condition~1 or conditions $1+2$. This ``quiescent time`` increases with $\tau_{\rm break}$ from $\sim 10\,{\rm Myrs}$ to over $50\,{\rm Myrs}$ in the case of condition 1. Imposing conditions $1+2$, the average time is $\sim10\,{\rm Myrs}$ in all cases. Note that this is an extremely short time scale in which the SFH variations must be aligned to produce the required properties.

Finally, the right panel of Figure~\ref{fig:uvj_model} shows two stochastic models with different statistics on the UVJ diagram. More stochasticity ({\it i.e.} smaller $\tau_{\rm break}$) results in larger scatter (color-coded points), overlapping with the observed color of \name. Longer correlation timescales results in a narrower parameter space (underlying gray points), similarly to a more simple burst$+$passive evolution model.

\begin{figure}[t!]
\centering
\includegraphics[angle=0,width=0.9\columnwidth]{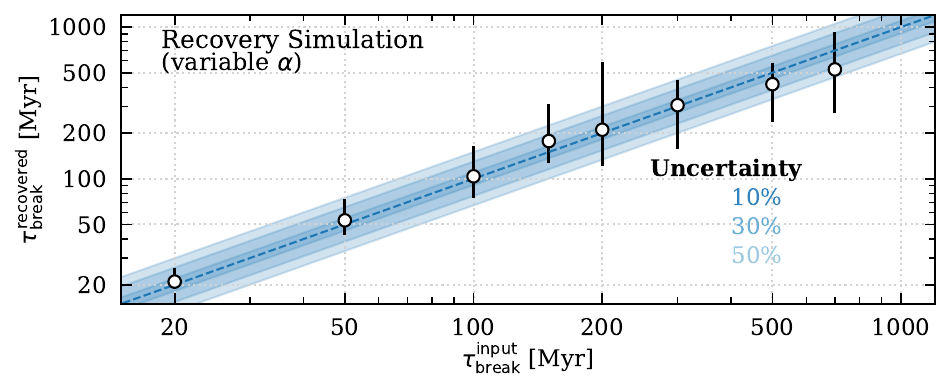}\\
\includegraphics[angle=0,width=1\columnwidth]{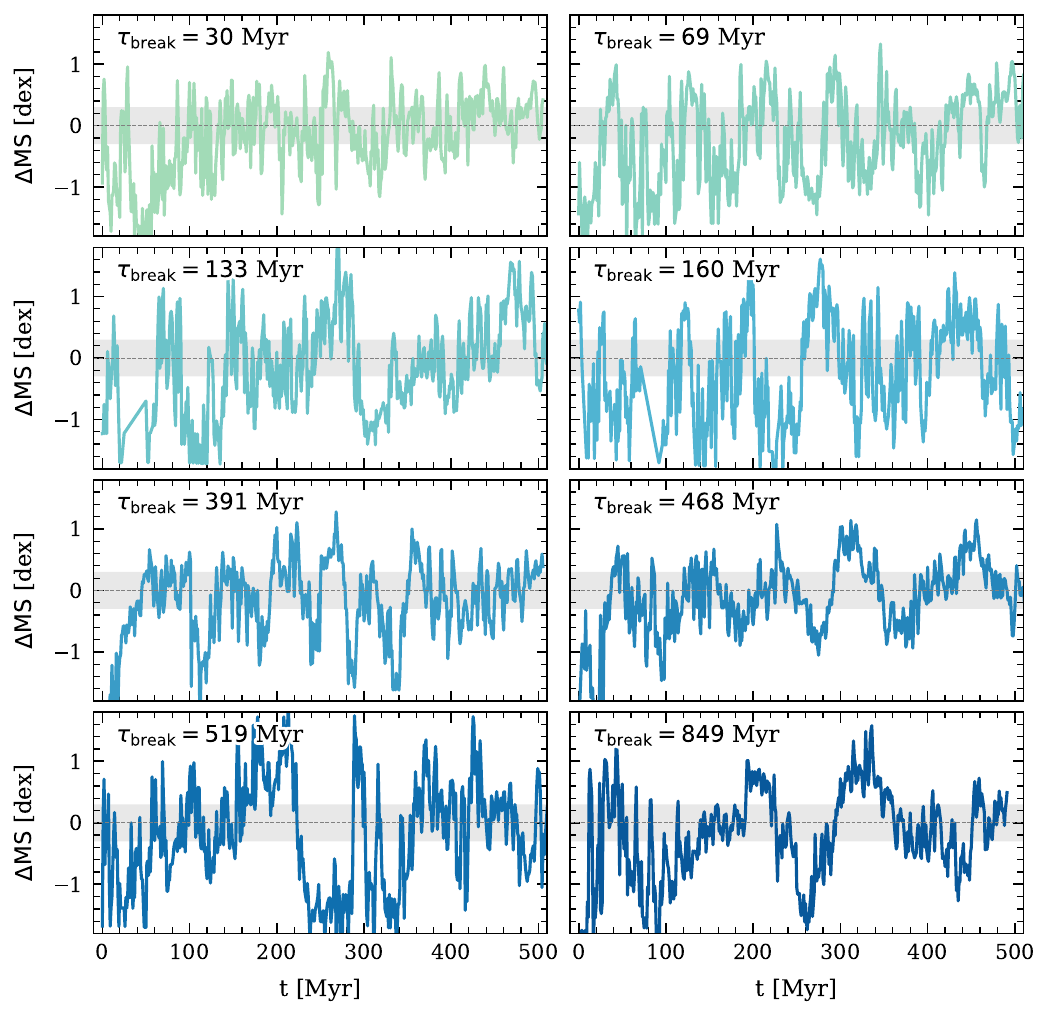}\\
\includegraphics[angle=0,width=0.95\columnwidth]{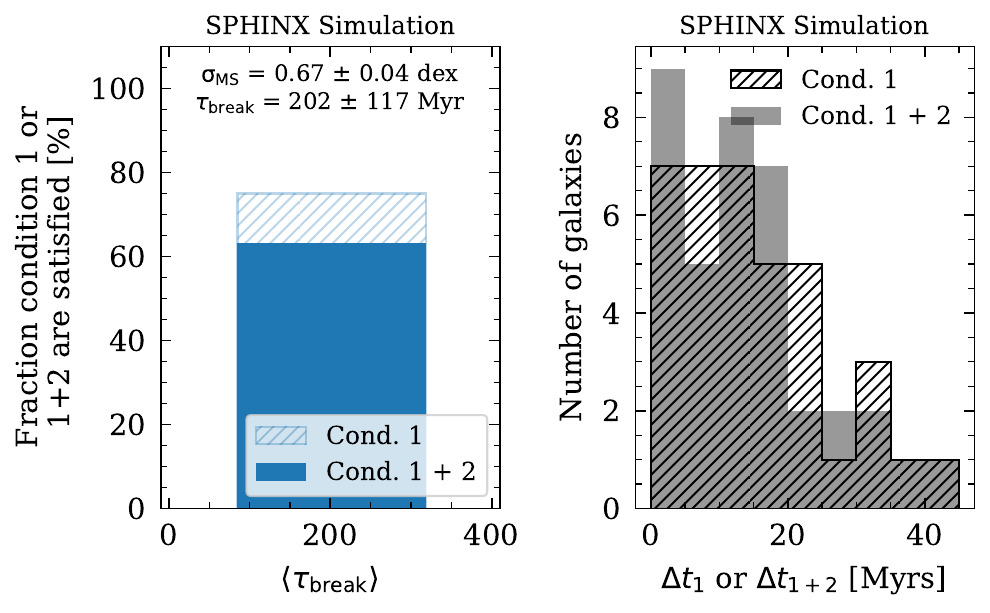}\vspace{-3mm}
\caption{{\it Top:} Recovery of $\tau_{\rm break}$ from simulated stochastic SFHs for variable $\alpha$. Uncertainties are shown in different shades of blue.
{\it Middle:} Examples \sphinxtwenty~SFHs from the $z=7$ slice across $500\,{\rm Myrs}$ normalized by the mean SFR (fit by 2-degree polynomial across the same time span) and sorted by fitted $\tau_{\rm break}$. The ``main sequence'' with $\pm0.3\,{\rm dex}$ scatter is indicated by the horizontal gray-shaded region. The galaxies are of similar stellar mass as \name.
{\it Bottom:} Fraction of the simulated galaxies going through observed conditions 1 (hatched) and both conditions $1+2$ (solid) over their lifetime until $z=7$ (left). Histogram (right) of corresponding average times of a galaxy spent in conditions 1 (black hatched) and $1+2$ (gray solid).
\label{fig:sphinxresults}}
\end{figure}

\begin{figure*}[t!]
\centering
\includegraphics[angle=0,width=0.96\columnwidth]{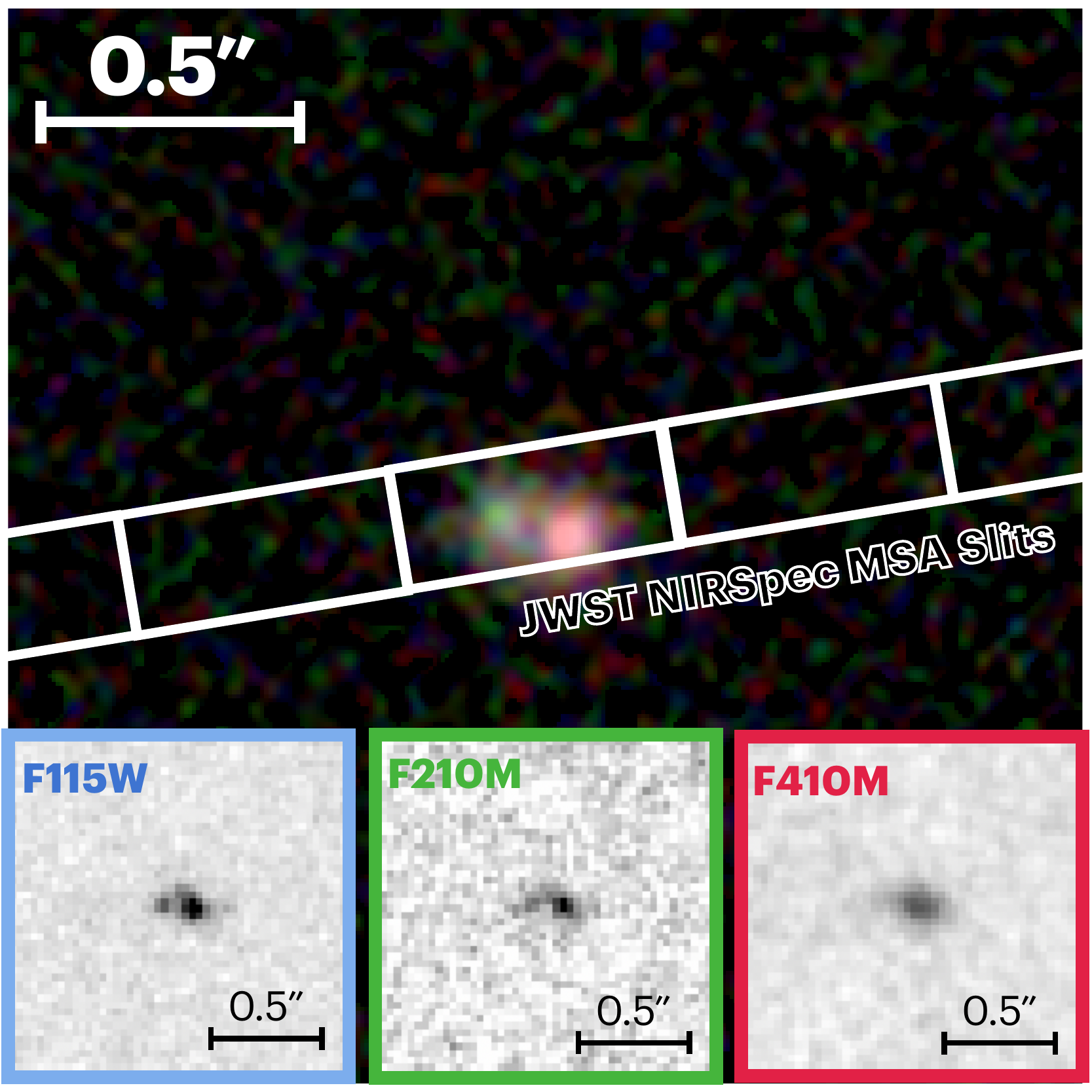}
\includegraphics[angle=0,width=1.12\columnwidth,trim=0mm 2.5mm 0 0]{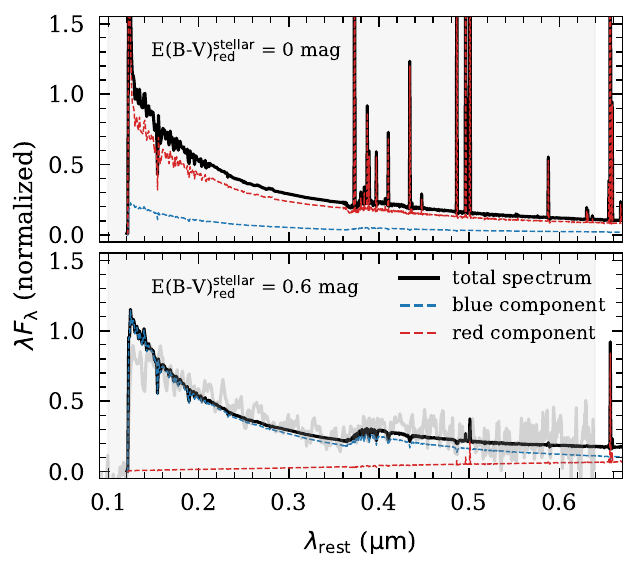}
\caption{{\it Left:} RGB image of \name~(MSA slit indicated). The image is generated with the \texttt{Galfit} models fit to the JWST/NIRCam F115w, F210M, and F410M images and convolved with the F115W PSF. There are indications of a blue and red component suggestive of different dust obscuration.
{\it Right:} Composite spectra of a blue (density bounded) and red (older, dust-obscured) component. The upper and lower panels show two different stellar continuum \ebmv~of the red component. A differential dust attenuation factor of $0.44$ is assumed. This shows that emission lines can be ``absent'' in the case of high dust attenuation in compact starbursts, while the blue continuum is dominated by the blue, dust-free young component of the galaxy (or galaxy merger). Even in the case of dust attenuation, \halpha~could be visible (however, outside of the NIRSpec wavelength coverage, gray region). The observed spectrum is shown in gray in the bottom panel.
\label{fig:ebmvvar}}
\end{figure*}

\subsubsection{Comparison to Simulations at $z=7$}\label{sec:sims}

In the following, we use more realistic SFH histories from simulations from the newly released \sphinxtwenty~cosmological radiation hydrodynamic simulations \citep{katz23}, which is the largest volume simulation ($20^3$ co-moving ${\rm Mpc^3}$ containing several thousands of star-forming galaxies at $z=6$) in the \sphinx~suite \citep{rosdahl18,rosdahl22}. The simulation models the large-scale process of cosmic reionization and the detailed physics of a multi-phase interstellar medium (ISM), and is therefore ideally suited for recovering the properties of galaxies at $z>6$. For more information on the details of this simulation, we refer to \citet{katz23}.

We first compare the more realistic SFHs of \sphinxtwenty~to our stochastic models. For this, we extract the SFHs of all $177$ galaxies (or halos) in the $z=7$ redshift slice. The galaxies have similar stellar masses as \name, hence are ideal for comparing. The SFHs are then normalized to their mean SFR (fit by a 2-degree polynomial) to obtain an estimate for $\sigma_{\rm MS}$. The resulting normalized SFHs are then fit with the PSD model described in Section~\ref{sec:stoch} to obtain a value for $\tau_{\rm break}$. The best-fit parameters for the $z=7$ \sphinxtwenty~population are $\sigma_{\rm MS} = 0.67\pm0.04\,{\rm dex}$, $\tau_{\rm break} = 202\pm117\,{\rm Myrs}$, and $\alpha=1.3\pm0.2$.
We verified the fitting method by simulating several $100$ SFHs for various $\tau_{\rm break}$ and $\alpha$ parameters, which we fit with the same methods. As shown in the top panel of Figure~\ref{fig:sphinxresults}, we are able to recover the true $\tau_{\rm break}$ to better than $50\%$ uncertainty independent of the input $\alpha$ value. The middle panels of Figure~\ref{fig:sphinxresults} shows examples of mean-normalized \sphinxtwenty~SFHs sorted by $\tau_{\rm break}$ obtained from our fits. 

The resulting combination of $\sigma_{\rm MS}$ and $\tau_{\rm break}$ suggests that we would be likely to identify a galaxy with similar properties as \name~in the \sphinxtwenty~volume. 
Indeed, we can calculate the probability in more detail, by simulating synthetic spectra from the SFHs of the \sphinxtwenty~galaxies directly, following the same recipe as outlined in Sections~\ref{sec:syntspec} and ~\ref{sec:occurrence}. Doing so suggests that about $60-80\%$ of the simulated galaxies go through similar observed properties as \name~ during their lifetime until $z=7$ (lower left panel, Figure~\ref{fig:sphinxresults}). The average ``quiescence time`` of galaxies with such observed properties is less than $20\,{\rm Myrs}$ on average (lower right panel).

Using these numbers from the \sphinxtwenty~simulation, we can estimate the fraction of galaxies, which we would expect to observe in a phase similar to \name. For this, we assume that $70\%$ of galaxies go through a temporary quiescent within $500\,{\rm Myrs}$, and for each of them, the quiescent phase occurs on average twice within that time and lasts on average for $10\,{\rm Myrs}$. We perform a Monte Carlo sampling approach with the above assumptions and evaluate the likelihood that a galaxy is observed in a quiescent phase at $z=7$ based on 1000 samplings\footnote{With the caveat that we assume that the galaxies lived for at least $500\,{\rm Myrs}$.}. Following this, we find a likelihood of $2.8\pm1.6\%$ to observe a galaxy similar to \name~at $z=7$. In other words, about $30^{+47}_{-13}$ star-forming galaxies at $5\times10^{8}$ need to be surveyed in order to find a temporarily quiescent galaxy. This is roughly consistent with the work by \citep{looser23b} who find one so-called (mini-)quenched galaxy ({\it i.e.} \name) among $\sim40$ star-forming galaxies in the mass range of $10^6-10^9\,{\rm M_\odot}$ at $5<z<11$.

\section{Impact of Dust Attenuation on Observed Quiescence}\label{sec:dust}

In the previous section, we have studied how a stochastic SFH can lead to the observed properties of a galaxy like \name. Abrupt changes in the SFH are likely the dominant reason \citep[see also][]{gelli23}, however, a blue$+$red composite spectrum could reconcile these observations as well, as reasoned below.

Several studies based on ALMA observations suggest a significant fraction of dust in galaxies even at $z\sim7$ \citep[e.g., ][]{faisst20,fudamoto20,fudamoto21,inami22,mitsuhashi23}.
Furthermore, thanks to large ALMA surveys such as ALPINE \citep{lefevre20,bethermin20,faisst20b} and REBELS \citep{bouwens22}, we have statistical evidence of spatial displacement between UV and far-IR emission in high-$z$ galaxies \citep[e.g.,][]{fudamoto21,jones20,bethermin23,devereaux23}. This suggests distinct regions of high and low dust obscuration. Similar conditions have been observed in quiescent galaxies at later times \citep[e.g.,][]{morishita22,lee24,setton24}.

In fact, there is observational evidence of a spatially variable dust attenuation in \name. The left panel of Figure~\ref{fig:ebmvvar} shows a color composite based on the JWST/NIRCam filters F115W, F210M, and F410M. To generate this image, we fit the two components (visible in the blue filters, see insets) using \texttt{Galfit} \citep{galfitpeng10} with positional priors based on F115W. The model images are then convolved by the F115W PSF and combined to an RGB image. The color-composite shows a blue component (east), separated from the main redder component (west). The color gradient could indicate spatially variable dust obscuration inside the galaxy or a blue merger component.

In the following, we study whether the spectrum of \name~could also be reconciled by a, possibly spatially unresolved, blue$+$red composite  as shown on the right panel of Figure~\ref{fig:ebmvvar}.
We model the blue component as a young ($40\,{\rm Myr}$), constant star-forming dust-free UV-emitting region that exhibits a high continuum escape fraction due to a density bounded nature. In such a setting, the ionizing radiation is escaping without producing significant nebular line emission \citep[e.g.,][]{rodriguez09,raiter10,topping22}. The search for such blue systems is ongoing with JWST and a few candidates have been found at $z>7$ (see \citealt{topping22} and references therein). The blue component would have a steep UV continuum with weak nebular emission.
In contrast, the red component is modeled as a post-starburst $60\,{\rm Myr}$ old dust-obscured region (maybe a post-starburst). Attenuation from dust generally impacts nebular emission more than the stellar continuum. This differential dust attenuation factor\footnote{$f~=~{\rm E(B-V)_{stellar}} / {\rm E(B-V)_{nebular}}$.} can reach $\frac{1}{f} = \frac{1}{0.44} \sim 2.3$ \citep[e.g.,][]{calzetti00,kashino13,reddy15}. Although $f$ is closer to unity at higher redshifts, a clumpy dust distribution can significantly alter this value and in the following we assume $f=0.44$ as worst case. Such a significantly dust obscured region could be devoid of emission lines (low $f$) and contributing little to the UV continuum.

The right panel of Figure~\ref{fig:ebmvvar} shows a possible resulting spectra. Without dust attenuation, the red component contributes significantly to the emission lines of the composite spectrum (upper panel). If \ebmv~=~0.6~mag (lower panel), the composite spectrum is practically devoid of emission lines. However, note that due to the redder wavelength of \halpha, this line should be detected (but in the case of \name~it is outside of the JWST/NIRSpec wavelength range). This situation resembles the spectrum of the massive quiescent galaxy GS-9209 found at $z=4.7$ devoid of strong optical emission lines except AGN-broadened \halpha~\citep{carnall23a}, with some important differences. First, GS-9209 is two orders of magnitude more massive, and second, the galaxy likely did not have any significant star formation within the past $100\,{\rm Myrs}$.

By using this model, we can predict JWST/MIRI fluxes in F560W and F770W. By convolving the blue$+$red component with \ebmv~=~0.6~mag (lower right panel in Figure~\ref{fig:ebmvvar}) with the corresponding filter curves, we find $m_{\rm F560W}=27.8$ and $m_{\rm F770W}=27.6$, respectively.

We note that generally the dust attenuation is expected to be low in low-mass high-redshift galaxies \citep[e.g.,][]{reddy23,sandles23}, hence an \ebmv~of 0.6\,mag is rare (but not unphysical as shown in the paragraph below). For example, based on dynamical mass measurements, the dust obscured galaxies found in \citet{fudamoto21} have stellar masses around $5\times10^{10}\,{\rm M_\odot}$ (although this may be an upper limit). Furthermore, the ALPINE galaxies are generally $>10^9\,{\rm M_\odot}$, thus at least a factor of two more massive than \name. Future observations with ALMA at rest-frame far-infrared wavelengths will be able to put solid constraints on its dust mass and obscured star formation. 

\begin{figure}[t!]
\centering
\includegraphics[angle=0,width=1.0\columnwidth]{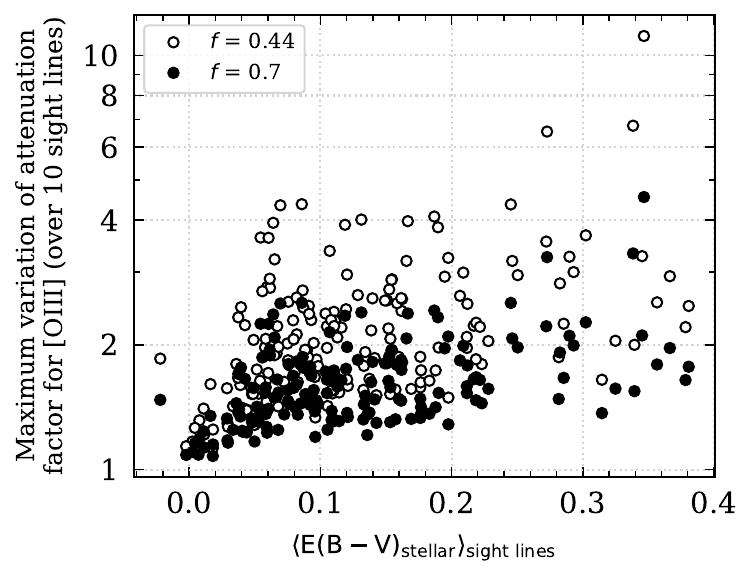}
\vspace{-6mm}
\caption{Maximum variation in attenuation factor of \oiii~among $10$ different sight lines as a function of median color excess of the stellar continuum in the \sphinxtwenty~simulation. The calculation assumes a \citet{calzetti00} starburst reddening curve and a differential dust attenuation factor of $f=0.44$ \citep[open symbols;][]{calzetti00} and $f=0.7$ \citep[closed symbols;][]{kashino13}. The extinction factor can change up to a factor of $10$ in certain cases (for $f=0.44$) due to geometric viewing angles, resulting in an important effect on the observation of nebular emission lines such as \oiii.
\label{fig:ebmvvarsim}}
\end{figure}

The \sphinxtwenty~simulation release provides a set of measurements along $10$ different sight lines for each of the simulated galaxies. We use the stellar continuum color excess \ebmv~measured along the different sight lines as an estimate of the patchiness of dust in these simulated $z=7$ galaxies. Specifically, we can study the impact dust attenuation on nebular emission lines by computing the attenuation factor ({\it i.e.} $10^{\frac{0.4}{f}\,{\rm E(B-V)}\,k^\prime(\lambda)}$) of \oiii~along these sight lines. For $k^\prime(\lambda)$ we assume a starburst reddening curve \citep{calzetti00}.
Figure~\ref{fig:ebmvvarsim} shows the variations of the \oiii~attenuation factor as a function of median \ebmv~(measured over all 10 sight lines) for two differential dust attenuation values of $f=0.44$ \citep{calzetti00} and $0.7$ \citep{kashino13}.
This test shows that the attenuation of nebular emission lines can change by factors of a few depending on the line of sight. In the most extreme case we find that the attenuation changes by a factor of $10$, which could have a significant impact on the detection of the line ({\it i.e.} an emission line at $10\sigma$ above the flux limit would be observed at only $1\sigma$).
Note that the abscissa in Figure~\ref{fig:ebmvvarsim} shows the stellar \ebmv~averaged over the $10$ sight lines. The maximum \ebmv~found on any sight line is $0.5\,{\rm mag}$, which is close to our assumed value of $0.6\,{\rm mag}$ in Figure~\ref{fig:ebmvvar}. Note that the simulated volume is rather small (177 galaxies at $z=7$), thus it is likely to reach these higher \ebmv~values.

The simulation therefore verifies the possibility of almost dust-free regions along with heavily dust-obscured patches. Overall, this shows that the spectrum of \name~can be naturally explained by a composition of a blue and red spatial component.

\section{Summary \& Conclusions} \label{sec:end}

The seemingly quiescent galaxy \name~at $z=7.3$ is a curious case. The combination of lacking emission lines and blue UV to optical ratio is difficult to reconcile in simulations and theoretical models. Specifically, a galaxy like \name~would not be classified as ``quiescent'' on common color diagrams (Figure~\ref{fig:uvj_model}), but rather as a galaxy with a short-term deficit of star formation.
We have explored the observational properties of galaxies such as \name~in relation to {\it (i)} a stochastic varying (bursty) star formation, and {\it (ii)} a spatially varying dust attenuation causing a blue and red composite spectrum.

The observed properties of \name~can be achieved shortly after a starburst (Figure~\ref{fig:uvj_model}). From exploring more realistic stochastic SFHs, we learned that a short star formation correlation time ($\tau_{\rm break} < 150\,{\rm Myr}$) is necessary to explain the galaxy's observed properties under the reasonable assumption of an average star formation oscillation amplitude around the mean of $\sigma_{\rm MS} = 0.5-0.6\,{\rm dex}$. This is consistent with recent measurements of the main-sequence scatter and star formation histories by JWST \citep{cole23}. Furthermore, a similar scatter may be necessary to reconcile the UV luminosity functions at $z>9$ \citep{shen23}. The average time spent in such a configuration is on the order of $10\,{\rm Myrs}$ or less (Figure~\ref{fig:stochastic}). A comparison to more realistic SFHs from the \sphinxtwenty~cosmological hydrodynamic simulation for similar-mass $z=7$ galaxies showed that about $60-80\%$ of simulated galaxies went through this state of temporary quiescence in the past and each phase of quiescence lasts about $10-20\,{\rm Myrs}$ (Figure~\ref{fig:sphinxresults}). Using these numbers, we estimated to find one temporarily quiescent galaxies among $\sim35$ star-forming galaxies at $z=7$, a number that is consistent with observational studies \citep{looser23b}.
However, we note that the \sphinxtwenty~simulation does not include black hole formation or AGN feedback due to its small volume \citep{katz23}. AGN may be important in the quenching of high-redshift galaxies at a similar mass range as \name~as shown both theoretically and observationally by recent studies \citep{xie24,juodzbalis23}.

In addition to a stochastic SFH, we explored the effect of spatially varying dust attenuation within a galaxy (or as part of a merging system) on the observed UV continuum and emission lines. 
We showed that a composite spectrum based on the combination of a UV-bright dust-free density bounded region and a red dust-obscured post-starburst region can result in a spectrum similar to the one of \name~(Figure~\ref{fig:ebmvvar}). Specifically, the density bounded region results in a steep UV continuum, while a high differential dust attenuation between stellar continuum and nebular emission of the red component leads to the suppression of nebular emission. Due to the lesser dust attenuation at longer wavelengths, we would still expect the galaxy to show \halpha~emission in this case (unfortunately not covered by current NIRSpec observations).
Based on the \sphinxtwenty~simulation set, we found that variations in the attenuation of the nebular lines due to dust along different sight lines can vary from factors of a few up to a factor of $10$ (in one case, Figure~\ref{fig:ebmvvarsim}). This is in agreement with UV and infrared emission offsets in recent ALMA observations. 
NIRCam imaging of \name~does reveal a blue and red component, which could therefore naturally explain the observed spectrum.

In summary, both stochastic SFH and dust obscuration can explain the spectrum of \name. Both cases are connected and are therefore not exclusive. Upcoming observations with ALMA of the far-infrared continuum and the \cii~line emission (\#2023.1.00521.S, PI: Faisst) will shed more light on the dust-obscured star formation of \name, which will help to constrain the above theories. 

\begin{acknowledgments}
{\it Acknowledgements:} We thank the referee for the insightful comments which significantly improved our original manuscript. Futhermore, we thank H. Katz for useful discussions about the \sphinxtwenty~simulation.
Some of the data presented in this article were obtained from the Mikulski Archive for Space Telescopes (MAST) at the Space Telescope Science Institute. The specific observations analyzed can be accessed via\dataset[10.17909/3c0d-5127]{https://doi.org/10.17909/3c0d-5127} and\dataset[10.17909/dw7f-zn13]{https://doi.org/10.17909/dw7f-zn13}.
\end{acknowledgments}

%

\vspace{5mm}
\facilities{JWST}


\software{
\texttt{astropy} \citep{astropy13,astropy18};
\texttt{DELightcurveSimulation} \citep{connolly16};
\texttt{FSPS} \citep{conroy09,conroy10};
\texttt{Galfit} \citep{galfitpeng10};
\texttt{Python-FSPS} \citep{johnson23}
          }





\bibliography{bibli}{}
\bibliographystyle{aasjournal}



\end{document}